\documentclass[11pt,superscriptaddress,aps,prd,preprint,showpacs,nofootinbib]{revtex4}
\usepackage[dvips]{graphicx}
\usepackage{amsfonts}
\usepackage{slashed}
\usepackage{xcolor}

\newcommand{\be}{\begin{equation}}
	\newcommand{\en}{\end{equation}}
\newcommand{\ba}{\begin{eqnarray}}
	\newcommand{\ea}{\end{eqnarray}}
\newcommand{\bea}{\begin{eqnarray}}
	\newcommand{\eea}{\end{eqnarray}}

\begin{document}

	\title{Radiative corrections in metric-affine bumblebee model}
	
	\author{Adria Delhom}
	\email[]{adria.delhom@uv.es}
	\affiliation{Departament de F\'{i}sica Te\`{o}rica and IFIC, Centro Mixto Universitat de
		Val\`{e}ncia - CSIC,\\
		Universitat de Val\`{e}ncia, Burjassot-46100, Val\`{e}ncia, Spain}
	
	\author{J. R. Nascimento}
	\email[]{jroberto@fisica.ufpb.br}
	\affiliation{Departamento de F\'{\i}sica, Universidade Federal da 
		Para\'{\i}ba,\\
		Caixa Postal 5008, 58051-970, Jo\~ao Pessoa, Para\'{\i}ba, Brazil}
	
	\author{Gonzalo J. Olmo}
	\email[]{gonzalo.olmo@uv.es}
	\affiliation{Departament de F\'{i}sica Te\`{o}rica and IFIC, Centro Mixto Universitat de
		Val\`{e}ncia - CSIC,\\
		Universitat de Val\`{e}ncia, Burjassot-46100, Val\`{e}ncia, Spain}

	\author{A. Yu. Petrov}
	\email[]{petrov@fisica.ufpb.br}
	\affiliation{Departamento de F\'{\i}sica, Universidade Federal da Para\'{\i}ba,\\
		Caixa Postal 5008, 58051-970, Jo\~ao Pessoa, Para\'{\i}ba, Brazil}
	
	\author{P. J. Porf\'{\i}rio}
\email[]{pporfirio@fisica.ufpb.br}
	\affiliation{Departamento de F\'{\i}sica, Universidade Federal da 
		Para\'{\i}ba,\\
		Caixa Postal 5008, 58051-970, Jo\~ao Pessoa, Para\'{\i}ba, Brazil}

	
	\begin{abstract}
		We consider the metric-affine formulation of bumblebee gravity, derive the field equations, and  show that the connection can be written as Levi-Civita of a disformally related metric in which the bumblebee field determines the disformal part. As a consequence, the bumblebee field gets coupled to all the other matter fields present in the theory, potentially leading to nontrivial phenomenological effects. To explore this issue we compute the weak-field limit and study the resulting effective theory. In this scenario, we couple scalar and spinorial matter to the effective metric which, besides the zeroth-order Minkowskian contribution, also has the vector field contributions of the bumblebee, and show that it is renormalizable at one-loop level. From our analysis it also follows that the non-metricity of this theory is determined by the gradient of the bumblebee field, and that it can acquire a vacuum expectation value due to the contribution of the bumblebee field. 
	\end{abstract}
	
	
	\maketitle

\section{Introduction}
\label{sec:intro}

The search for signatures of Lorentz symmetry breaking (LSB) has called a great deal { of} attention over the last decades. {String theories became the first motivation for} the studies of LSB \cite{Kostelecky:1989jp, Kostelecky:1989jw, Kostelecky:1991ak, Kostelecky:1994rn}. Non-perturbative string/M-theory predict extended objects, for example, D-branes, which spontaneously break the Lorentz and translation symmetries of the bulk spacetime \cite{Gliozzi:2011hj}. Other motivations for Lorentz symmetry breaking in other quantum gravity contexts such as {\it e.g.} loop quantum gravity, quantum foam, etc. have also been exploited in the literature \cite{Gambini:1998it, Calcagni:2016zqv, Bojowald:2004bb, Alfaro:2004aa, AmelinoCamelia:2001dy}. Given the experimental difficulties to probe such theories at the Planck scale, it is useful to follow a bottom-up approach and consider effective field theories (EFT) containing operators which are a source for local Lorentz violation (LV). The Standard-Model Extension (SME) proposed in \cite{Colladay:1996iz, Colladay:1998fq}, as well as its { generalization} to a gravitational sector described by GR at low energies \cite{Kostelecky:2003fs}, is a general framework which incorporates all possible Lorentz and CPT-violating operators that can be written in terms of fields which belong to some representation of the Poincar\'e group\footnote{Note that, if Lorentz symmetry is explicitly broken at some scale, the physical (asymptotic) states will not in general be described by representations of the Poincar\'e group, and therefore the most general low-energy effective theory for LV could contain operators which cannot be written in terms of fields that belong to some representation of the Poincar\'e group.}. The operators that appear in these effective field theories could be understood as contributions to the low-energy effective action of some UV theory with a spontaneous breakdown of Lorentz symmetry due to some set of non-scalar fields that develop a non-trivial vacuum expectation value (VEV).  In the presence of gravity, however, forcing the VEVs to be constant would imply constraints on the allowed background space-time for consistency of the model, as happens for instance in Einstein-aether gravity \cite{Jacobson:2000xp} or in GR with a Chern-Simons term \cite{Jackiw:2003pm}. Allowing these Higgsed fields to be dynamical guarantees the general covariance of the theory as a consequence of them satisfying their respective field equations \cite{Jacobson:2000xp}. 

Among the several models that have been used to describe the gravitational sector of the SME, the bumblebee model \cite{Kostelecky:2003fs,Bluhm,Seifert:2009gi, Nascimento:2014vva, Casana:2017jkc}  can be singled out from an EFT perspective as it involves only first-order curvature terms in the Lagrangian, which are the dominant ones at low curvatures (energies). In this model, a vector field driven by a non-trivial potential which spontaneously breaks Lorentz symmetry is non-minimally coupled to the gravitational sector through the Ricci tensor. Other extensions including higher-order curvature couplings which would be relevant when considering strong-gravitational field regimes have also been considered within the SME framework.

In this context, one must emphasize the fundamental difference between the bumblebee gravity and other known vector-tensor gravity theories such as generalized Proca models discussed f.e. in \cite{GenPr1,GenPr2,GenPr3} and so-called Proca-Nuevo ones, considered f.e. in \cite{ProNu1,ProNu2,ProNu3}. Unlike these theories, characterized by the presence of additional derivative-dependent terms in the vector sector and actually representing themselves as galileon-like theories, the bumblebee model, either in flat or in a curved space-time, includes a new potential term that allows for spontaneous Lorentz symmetry breaking, which is not observed in Proca-like theories.

The large majority of works in the available literature have treated the LV gravitational models in the metric approach, where the geometry is {\it a priori} assumed to be fully described only by the metric tensor, this being the only dynamical geometrical object. In contrast, one could employ the metric-affine approach in which no {\it a priori} relation between the metric and the affine connection is assumed. The fact that the metric approach is more popular than the metric-affine (or Palatini) one can be related to the fact that the current paradigm of gravitation as a geometric phenomenon initiated by Einstein arose in a time in which the only known geometries were of Riemannian type. Mathematicians soon realized that more general geometries were possible, and a first step in this direction was to relax the compatibility condition between metric and affine connection, thus giving rise to metric-affine geometry \cite{Eisenhart-1927}. The metric-affine approach must thus be seen as a reduction on the number of assumptions and, therefore, as potentially providing a more general phenomenological scenario than the metric approach. Note, in this sense, that whether the space-time geometry is closer to a Riemannian type, to a metric-affine type, or to something else is a question to be answered by experiments, not by conventions \cite{Feigl_etal-1953}.  To our knowledge, the first works that considered LSB in generic metric-affine geometries studied the role that could be played by generic torsion \cite{Kostelecky:2003fs} and non-metricity \cite{Foster:2016uui} backgrounds. Other non-Riemannian geometries such as Riemann-Finsler space-times have also been considered in the context of LSB \cite{Kostelecky:2011qz}. These works paved the road to study particular examples in which the non-trivial non-metricity or torsion VEV that spontaneously breaks the Lorentz symmetry is generated by the gravitational dynamics. In this direction, the metric-affine version of the bumblebee model has recently been formulated in \cite{Delhom:2019gxg}. That work focused on the classical aspects of the theory, showing that this model can be embedded in a generalization of the Ricci-based-gravity (RBG) class \cite{BeltranJimenez:2017doy,Afonso:2018bpv,Afonso:2018mxn,Delhom:2019zrb,BeltranJimenez:2019acz,Jimenez:2020dpn,Jimenez:2020iok,Latorre:2017uve,Delhom:2019wir,BeltranJimenez:2021iqs} where non-minimal couplings between some matter fields and the Ricci tensor are allowed. This allowed to find the solution to the connection equations with the same procedure as in minimally coupled RBGs, as well as to show that the theory admits an Einstein-frame representation where a tower of effective LV operators appear in the matter sector due to the bumblebee VEV and its non-minimal coupling to the Ricci tensor. In that work, it was also shown that, due to the form of the non-minimal coupling and the absence of higher-order curvature terms, the non-metricity tensor is completely determined by the (covariant) derivatives of the bumblebee field of the model. Finally, the stability conditions of the classical effective theory in the weak gravitational field limit and coupled to a scalar and a Dirac field were considered. 

A natural step to follow, and the focus of this work, is to understand the role played by quantum corrections in the LV coefficients that arise in the classical theory. In this sense, it is well established that purely quantum effects can account for symmetry breaking \cite{Coleman} (a phenomenon known as dynamical symmetry breaking). It is thus very important to determine if perturbative corrections within the bumblebee gravity may contribute to the effective potential and allow for dynamical Lorentz symmetry breaking, see f.e. \cite{Gomes:2007mq,AltKost}. In the current literature on bumblebee gravity, only tree-level studies of the spontaneous LV have been performed in cosmological \cite{Maluf1,Maluf2} and black hole scenarios \cite{Maluf3}, where the role of privileged space-time directions is played by time and radial coordinates. The study of generic quantum corrections in bumblebee gravity, either coupled to some matter or not, is thus an open problem. Moreover, since all considerations of spontaneous Lorentz symmetry breaking, including the papers cited above, were performed within the metric formalism, it becomes necessary to extend these studies to the metric-affine framework in order to identify characteristic features that may help tell both formalisms apart or that may favor one approach over the other. In this paper, we calculate these perturbative corrections. Explicitly, we will compute the one-loop corrections to the classical effective action in the weak gravitational field limit for a non-trivial bumblebee background, paying special attention to the contribution of the non-minimal couplings between the bumblebee field and the matter sources to the one-loop quantum corrections. To that end, we will consider quantum Dirac fields as the matter content of the model. As explained in \cite{Delhom:2019gxg}, these will be described by a classical effective Lagrangian which can be expanded in powers of the non-minimal coupling parameter $\xi$. Given that the terms of order $\xi^2$ would be of the same order as quadratic curvature corrections which are not present in the bumblebee model, we will stick to $\mathcal{O}(\xi)$ corrections for consistency.

The paper is organized as follows. In section \ref{sec2} we review the main features of the classical metric-affine bumblebee model. In section \ref{sec3} we study the quantum corrections at the one-loop level in the weak gravitational field limit and up to first order in $\xi$. Finally, we conclude in section \ref{sec4}. Throughout the paper we use natural units $c=\hbar=1$.

\section{The metric-affine bumblebee model}\label{sec2}

We start this section by writing down the metric-affine bumblebee action in curved space-time \cite{Delhom:2019gxg}
\begin{eqnarray}
	\nonumber S_{B}&=&\int d^4 x\,\sqrt{-g}\Big[\frac{1}{2\kappa^2}\Big(R(\Gamma)+\xi B^{\alpha} B^{\beta} R_{\alpha\beta}(\Gamma)\Big)-\frac{1}{4}B^{\mu\nu}B_{\mu\nu}-V(B^{\mu}B_{\mu}\mp b^2)\Big] +\\
	&+& \int d^4 x\,\sqrt{-g}\mathcal{L}_{M}(g_{\mu\nu},\psi),
	\label{bumblebee}
\end{eqnarray} 
where $R(\Gamma)$ is the Ricci scalar, $R_{\mu\nu}(\Gamma)$ is the Ricci tensor, $\mathcal{L}_{M}$ is the Lagrangian of the matter sources, $B_{\mu}$ is the bumblebee field, $B_{\mu\nu}=2\partial_{[\mu}B_{\nu]}$ its field strength, $V$ denotes the bumblebee potential, and $\kappa^2=8\pi G$. Here, $\xi$ is a small constant characterizing the inverse square of some large energy scale. The action of the model is defined {within} the metric-affine approach, which means that the connection is not assumed to bear any {\it a priori} relation with the metric.  The potential $V$ is such that the vector field $B_{\mu}$ acquires a non-zero VEV, say $B_{\mu}=b_{\mu}$ with $b^2>0$, which triggers a (local) spontaneous breaking of Lorentz symmetry. Then, at low energies, the effects of a preferred frame will in principle permeate observables which are related to the bumblebee field. Note that the antisymmetric part of the Ricci tensor is irrelevant, which guarantees that the model satisfies a projective symmetry. This feature has been seen to be relevant in order to avoid the propagation of ghost-like degrees of freedom associated to the affine connection \cite{BeltranJimenez:2019acz,Jimenez:2020dpn,Aoki:2019rvi}.  Regarding the matter Lagrangian, we assume that the matter fields are minimally coupled to the metric and that there is no explicit coupling to the connection. As we will see, this facilitates the resolution of the connection equation and, in particular, of its torsional part.\footnote{In the case of fermions, a coupling to the axial part of the torsion may be naturally included. This generates a contribution to the hypermomentum in the connection equation but does not have a critical impact on the symmetric part of the connection, which is the relevant piece in our analysis.} 

After some algebra, the field equations obtained upon varying (\ref{bumblebee}) with respect to metric, connection, and bumblebee field are respectively
\begin{eqnarray}
	\kappa^2  T_{\mu\nu} &=&G_{(\mu\nu)}(\Gamma)-\frac{\xi}{2}g_{\mu\nu}\bigg( B^{\alpha}B^{\beta}R_{\alpha\beta}(\Gamma)\bigg)
	+2\xi\bigg(B_{(\mu}R_{\nu)\beta}(\Gamma)\bigg)B^{\beta}, \label{Riccieq}\\
	0&=&\nabla_{\lambda}^{(\Gamma)}\bigg[\sqrt{-g}g^{\mu\alpha}\bigg(\delta^{\nu}_{\alpha}+\xi B^{\nu}B_{\alpha}\bigg)\bigg],\label{connectioneq}\\
	\nabla_{\mu}^{(g)}B^{\mu\nu}&=&-\frac{\xi}{\kappa^2}g^{\nu\alpha}B^{\beta}R_{\alpha\beta}(\Gamma)+2 V^{\prime}B^{\nu},\label{bumblebeeeq}
	\label{PDE}
\end{eqnarray}
where $T_{\mu\nu}=T_{\mu\nu}^{M}+T_{\mu\nu}^{B}$ encodes the contributions of the matter sources $T_{\mu\nu}^{M}$ as well as the contribution of the bumblebee field which does not couple to the curvature, namely
\begin{eqnarray}
	T_{\mu\nu}^{M}&=&-\frac{2}{\sqrt{-g}}\frac{\delta(\sqrt{-g}\mathcal{L}_{M})}{\delta g^{\mu\nu}},\\
	T_{\mu\nu}^{B}&=& B_{\mu\sigma}B_{\nu}^{\ \sigma}-\frac{1}{4}g_{\mu\nu}B^{\alpha}_{\ \sigma}B^{\sigma}_{\ \alpha}-V g_{\mu\nu}+2V^{\prime}B_{\mu}B_{\nu}.
\end{eqnarray}
Notice that the connection equation is algebraic, which implies that it is an auxiliary field which does not propagate new degrees of freedom and can be integrated out from the action if one finds the solution to (\ref{connectioneq}). The solution to the connection equation is given by the Levi-Civita connection of a new metric $h^{\mu\nu}$ defined by
\begin{equation}
	h^{\mu\nu}=\frac{1}{\sqrt{1+\xi X}}(g^{\mu\nu}+\xi B^{\mu}B^{\nu}),\label{hmn}
\end{equation}  
where $X\equiv g^{\mu\nu}B_{\mu}B_{\nu}$. By performing appropriate field redefinitions and integrating out the connection, the action (\ref{bumblebee}) can be recast in an Einstein-Hilbert-like form as
\begin{eqnarray}\label{EinsteinBumblebee}
	\tilde{\mathcal{S}}_{BEF}&=&\int d^{4} x \sqrt{-h}\frac{1}{2 \kappa^{2}} R(h)+\overline{\mathcal{S}}_{m}\left(h_{\mu \nu}, B_{\mu}, \psi\right)\;.
\end{eqnarray} 
Notice that the above action (\ref{EinsteinBumblebee}) is the Einstein-Hilbert action for the new metric $h^{\mu\nu}$ coupled to a modified matter sector $\bar{\mathcal{S}}_{\rm m}$, which encodes a tower of non-linear interactions between the bumblebee field and all the fields in the matter sector $\mathcal{S}_{\rm m}$ appearing in the original action (\ref{bumblebee}). The interested reader is referred to \cite{Delhom:2019gxg} for details on the physical equivalence between (\ref{bumblebee}) and (\ref{EinsteinBumblebee}). Note that the representation (\ref{EinsteinBumblebee}) makes it clear that in this theory gravitational waves propagate on the light cone defined by the metric $h_{\mu\nu}$. One could thus observe subluminal/superluminal propagation of the tensorial perturbations of $g_{\mu\nu}$ (the physical gravitational waves, as $g_{\mu\nu}$ is the metric minimally coupled to the standard model matter) in regions of very intense bumblebee field. A detailed quantitative analysis would be necessary in order to determine if current data could be used to constrain this model with gravitational wave data, but that lies beyond the scope of this work and will be considered elsewhere. 

From the connection equation it is clear that, generally, the VEV of the bumblebee field provides a VEV for the non-metricity tensor $Q_{\mu}{}^{\alpha\beta}\equiv\nabla_\mu g^{\alpha\beta}$, realizing dynamically the idea presented in \cite{Foster:2016uui} of having a non-trivial non-metricity background that induces spontaneous breaking of Lorentz symmetry. Let us now quickly comment on the equations of motion for the bumblebee field. Either substituting (\ref{hmn}) into (\ref{bumblebeeeq}) or directly deriving the equation from the Einstein-frame action (\ref{EinsteinBumblebee}), the bumblebee is described by an effective Proca-like equation 
\begin{equation}
	\nabla_{\nu}^{(g)} B^{\nu\mu}=\mathcal{M}^{\mu}_{\,\,\nu}B^{\nu},
	\label{Proca}
\end{equation}
where the derivative operator $\nabla_{\nu}^{(g)}$ denotes covariant derivative with respect to the Levi-Civita connection of the metric $g_{\mu\nu}$, and the effective  mass term $\mathcal{M}^{\mu}_{\,\,\nu}$ is given by
\begin{eqnarray}
	\nonumber\mathcal{M}^{\mu}_{\,\,\nu}&=&\left(\frac{\xi T}{2+3\xi X}+\frac{\xi^2 B^{\alpha}B^{\beta}T_{\alpha\beta}}{(1+\xi X)(2+3\xi X)}+2V^{\prime}\right)\delta^{\mu}_{\,\,\nu}-\\
	&-&\frac{\xi}{(1+\xi X)}T^{\mu\alpha}g_{\nu\alpha} \ .
	\label{masseff}
\end{eqnarray}
This object $\mathcal{M}^{\mu}_{\,\,\nu}$ describes a mass term (from the $2V'$ term) as well as self interactions and interactions with all the other fields in the matter sector through the stress-energy tensor. Notice that the stability of the above equation is not guaranteed because the determinant of $\mathcal{M}^{\mu}_{\,\,\nu}$ may not be positive definite. 

 Before concluding this section,  we would like to note that the field equations obtained above in the metric-affine formulation are quite different from those derived in a purely metric approach (see,  for instance,  \cite{Maluf3}).   In the latter case,  the metric variation involves an integration by parts due to the {\it a priori} imposed compatibility between metric and connection that induces derivatives of the bumblebee on the right hand side of (\ref{Riccieq}).  Those derivatives will then appear in the bumblebee equation via the contraction $B_{\alpha}R^{\alpha\nu}$,  thus leading to a differential structure different from our Eq. (\ref{Proca}).  In addition,  in the metric-affine formulation, the structure of the action (\ref{EinsteinBumblebee}) in the Einstein-frame variables puts forward that $h_{\mu\nu}$ is sensitive to the total amounts of mass and energy,  which implies that the weak field limit is achieved when $h_{\mu\nu}\approx \eta_{\mu\nu}$.  As a result,  all couplings to $g_{\mu\nu}$ involve the presence of the scalar $X$ and of the disformal term $B_\mu B_\nu$ (see (\ref{hmn})),  which lead to new interactions between the bumblebee and all the other matter sources in ways that do not appear in the metric formulation.  Hence,  formally,  one concludes that metric and metric-affine formulations of the bumblebee model lead to completely different theories.

\section{Radiative corrections}\label{sec3}

In this section we focus our attention on the quantum dynamics of spinor matter fields in the weak gravitational field limit, where $h_{\mu\nu}\approx \eta_{\mu\nu}$ or equivalently, neglecting $\mathcal{O}(\xi^2)$ terms, $g_{\mu\nu}\approx\eta_{\mu\nu} + \xi(B_{\mu}B_{\nu}-\frac{1}{2}X\eta_{\mu\nu})$, see \cite{Delhom:2019gxg}.  To that end, we have to specify a particular form for the bumblebee potential, which we will choose to be the usual Mexican hat potential, given by
\begin{equation}
	V(B^\mu B_\mu\mp b^2)=\frac{\lambda}{4}\big(B^\mu B_\mu\mp b^2\big)^2,
\end{equation}
where $\lambda$ is a positive weak coupling. Here $b^2>0$ and the $\mp$ sign accounts for the possibility of having a space- or time-like bumblebee VEV respectively. Notice that, after the field redefinition of $g_{\mu\nu}$ in terms of $h_{\mu\nu}$ and $B_\mu$ is implemented, new effective self couplings will arise in the bumblebee action.  In this scenario the spinor and bumblebee actions in the Einstein-frame read
\begin{eqnarray}
	\nonumber\label{ap2}\mathcal{L}_{spEF}&=&\bar{\Psi}\left(i\gamma^{\mu}\partial_{\mu}-m\right)\Psi-\frac{i}{4}\xi\bar{\Psi}B^{\nu}B_{\nu}\gamma^{\mu}\partial_{\mu}\Psi-
	\frac{i}{2}\xi\bar{\Psi}\left(B^{\nu}\gamma_{\nu}\right)B^{\mu}\partial_{\mu}\Psi+\frac{\xi}{2} m B^{\nu}B_{\nu} \bar{\Psi}\Psi-\\
	\nonumber&-&i\frac{\xi}{4}\bar{\Psi}\Big(B_{\alpha}\left(\partial_{\mu} B^{\alpha}\right)+B^{\nu}\left(\partial_{\nu} B_{\mu}\right)+
	(\partial_\alpha B^\alpha) B_{\mu}\Big)\gamma^{\mu}\Psi+\mathcal{O}(\xi^2),\label{SpinLagPert}\\
	\nonumber{\cal L}_{BEF}&=&-\frac{1}{4}B_{\mu\nu}B^{\mu\nu}+\frac{M^2}{2}B^2-\frac{\Lambda}{4}(B^2)^2+\\ \nonumber&+&\frac{\xi}{2}\Big[B^{\mu\nu}B^\alpha{}_{\nu}B_\mu B_\alpha-\frac{1}{4}B_{\mu\nu}B^{\mu\nu}B^2-\frac{3}{4}\Lambda(B^2)^3\Big]+
	\mathcal{O}(\xi^2)
	\label{BumbLagPert}
\end{eqnarray} 
respectively, where the bumblebee effective mass is given by
$M^2= \lambda b^2(\pm 1+\frac{\xi}{4}b^2)$ and $\Lambda\equiv \lambda(1\pm 2\xi b^2)$. From here on, indices are raised and lowered with the Minkowski metric, so that $B^2=\eta_{\mu\nu}B^\mu B^\nu$ and so on, which is consistent with neglecting $\mathcal{O}(\xi^2)$ terms. As explained in \cite{Delhom:2019gxg}, the new interaction terms arise due to the non-minimal coupling between the bumblebee and the affine connection through the Ricci tensor. Note that all aforementioned non-linear interactions are not expected to emerge in the metric approach because there is no direct coupling between the bumblebee and the matter sources, as we already discussed at the end of the previous section. Note also that no triple vertices arise by construction. 
Here we assume that the spontaneous Lorentz symmetry breaking is generated by the potential, in our case looking like $V=-\frac{M^2}{2}B^2+\frac{\Lambda}{4}(B^2)^2$, since higher-order terms coming from the non-metricity contributions, can be suppressed by different degrees of $\xi$. This potential has been discussed in great detail at the tree level in \cite{Bluhm:2008yt}, where the corresponding classification of potentials is presented.


Following \cite{Gomes:2007mq}, let us now compute the quantum corrections on top of a stable non-trivial bumblebee VEV characterized by $<B^\mu>=\beta^\mu$, where {$\beta^2=\pm b^2$}. The dynamics of small perturbations of the bumblebee field can then be analyzed by expanding the Lagrangian (\ref{BumbLagPert}) around the vacuum as $B_{\mu}=\beta_{\mu}+\tilde{B}_{\mu}$. This leads to the following Lagrangian for the perturbations
\begin{eqnarray}
\nonumber{\cal L}_{BEF}^{\rm pert}&=&-\frac{1}{4}\tilde{\eta}_{\mu\alpha}\tilde{B}^{\mu}_{\,\,\nu}\tilde{B}^{\alpha\nu}-\Lambda(\beta_\mu\tilde{B}^\mu)^2-\frac{\Lambda}{4}(\tilde{B}^2)^2+\\
\nonumber&+&\xi\Big[\frac{1}{2}\tilde{B}^{\mu\nu}\tilde{B}^\alpha{}_\nu\tilde{B}_\alpha\tilde{B}_{\mu}-\frac{1}{4}\tilde{B}^{\mu\nu}\tilde{B}_{\mu\nu}(\beta_\rho\tilde{B}^\rho)+\\
&+&\tilde{B}^{\mu\nu}\tilde{B}^\alpha{}_\nu\tilde{B}_\alpha\beta_{\mu}-\frac{1}{8}\tilde{B}^{\mu\nu}\tilde{B}_{\mu\nu}\tilde{B}^2\Big]+\mathcal{O}(\xi^2,\lambda\xi),
\label{BumbLagVEV}
\end{eqnarray}
where the kinetic term for the perturbations interacts with the background by coupling to the effective metric 
\begin{eqnarray}
\tilde{\eta}_{\mu\nu}\equiv\eta_{\mu\nu}\left(1+\frac{\xi\beta^2}{2}\right)-2\xi \beta_{\mu}\beta_{\nu}.
\end{eqnarray}
We note that in this case, due to the non-trivial minimum, the Maxwell-like term is rescaled looking like $-\frac{1}{4}\tilde{B}_{\mu\nu}\tilde{B}^{\mu\nu}(1+\frac{\xi\beta^2}{2})$, and the aether-like term arises. 

The free (linearized) equations of motion for the vector field are
\begin{eqnarray}
\nonumber 0&=&\partial_{\mu}\tilde{B}^{\mu\nu}\left(1+\frac{\xi\beta^2}{2}\right)-\xi \beta_{\mu}\beta_{\alpha}\partial^{\mu}\tilde{B}^{\alpha\nu}+\xi \beta^{\nu}\beta_{\alpha}\partial_{\mu}\tilde{B}^{\alpha\mu}-
2\Lambda\beta^{\nu}\beta_{\alpha}\tilde{B}^{\alpha},
\end{eqnarray}
or, in terms of the effective metric, they look like
\begin{equation}
	\tilde{\eta}_{\mu[\nu}\partial^{\nu}\tilde{B}^{\mu}_{\,\,\alpha]}+M_{\alpha\mu}\tilde{B}^{\mu}=0,
\end{equation}
where $M_{\alpha\mu}=-2\Lambda\beta_{\alpha}\beta_{\mu}$ is the effective mass-squared tensor.
Taking the divergence of this equation which must be zero, we find that, 
instead of the usual condition $(\partial\cdot B)=0$, we will have an essentially new condition $(\beta\cdot\partial)(\beta\cdot \tilde{B})=0$ (from the formal viewpoint, this condition is explained by the fact that our mass term is also aether-like, without the usual Proca mass term).
Under this condition, the free action of the vector field becomes
\begin{eqnarray}
{\cal L}&=&-\frac{1}{4}\tilde{B}_{\mu\nu}\tilde{B}^{\mu\nu}\left(1+\frac{\xi\beta^2}{2}\right)-\frac{\xi}{2}\tilde{B}_{\mu}[\Box\beta^{\mu}\beta^{\nu}+
(\beta\cdot\partial)^2\eta^{\mu\nu}]\tilde{B}_{\nu}-\Lambda(\beta^{\alpha}\tilde{B}_{\alpha})^2.
\end{eqnarray}

Now, let us define the vector-spinor interaction vertices for the non-zero vacuum and calculate the contributions to the two-point function (the four-point function can be calculated along the same lines but the computation is much more cumbersome for such a vacuum). {Substituting} $B_{\mu}=\tilde{B}_{\mu}+\beta_{\mu}$ into the EF spinor Lagrangian (\ref{SpinLagPert}) we arrive at
\begin{eqnarray}
\nonumber{\cal L}_{sp}&=&\bar{\Psi}\bigg[i\gamma^{\mu}\partial_{\mu}-m\left(1-\frac{1}{2}\xi\beta^{2}\right)-
\frac{i}{2}\xi\left(\beta^{\mu}\beta^{\nu}+\frac{1}{2}\beta^2\eta^{\mu\nu}\right)\gamma_{\mu}\partial_{\nu}\bigg]\Psi-\nonumber\\
\nonumber&-&\bar{\Psi}\bigg[\frac{i}{2}\xi\big(\tilde{B}^{\mu}\beta^{\nu}+\tilde{B}^{\nu}\beta^{\mu}+
(\tilde{B}\cdot\beta)\eta^{\mu\nu}\big)\gamma_{\mu}\partial_{\nu}-\xi m\tilde{B}^{\nu}\beta_{\nu}\bigg]\Psi-\nonumber\\
&-&\bar{\Psi}\left[\frac{i}{2}\xi\left(\tilde{B}^{\mu}\tilde{B}^{\nu}+\frac{1}{2}\tilde{B}^2\eta^{\mu\nu}\right)\gamma_{\mu}\partial_{\nu}-\frac{1}{2}\xi m\tilde{B}^2\right]\Psi-\nonumber\\
\nonumber&-&\frac{i}{4}\xi\bar{\Psi}\bigg[\tilde{B}_{\alpha}(\partial_{\mu} \tilde{B}^{\alpha})+\beta_{\alpha}(\partial_{\mu} \tilde{B}^{\alpha})+\tilde{B}^{\nu}(\partial_{\nu} \tilde{B}_{\mu})+\\
&+&\beta^{\nu}(\partial_{\nu} \tilde{B}_{\mu})+(\partial_\alpha \tilde{B}^\alpha) \tilde{B}_{\mu}+(\partial_\alpha \tilde{B}^\alpha) \beta_{\mu} \bigg]\gamma^{\mu}\Psi.\label{SpinLagVEV}
\end{eqnarray}
We assume $\xi$ to be small (recall that it has dimensions of inverse mass squared, so that it can be treated as an inverse square of some large mass scale). So, we can consider corrections of first order in $\xi$ only, which are given by one-vertex graphs and yield contributions to the two-point function. This actually means that we need only quartic vertices. We consider the graphs given in Fig. 1 which only yield contributions to the two-point function. To calculate those graphs, it remains to write down the background-dependent propagators. We have
\begin{eqnarray}
<\psi(-k)\bar{\psi}(k)>&=&i\bigg[\gamma^{\mu}k_{\mu}-m\left(1-\frac{1}{2}\xi\beta^2\right)-\frac{1}{2}\xi(\beta^{\mu}\beta^{\nu}+
\frac{1}{2}\beta^2\eta^{\mu\nu})\gamma_{\mu}k_{\nu}\bigg]^{-1};\\
\nonumber<\tilde{B}_{\alpha}(-k)\tilde{B}_{\beta}(k)>&=&
i\bigg[(-k^2\eta^{\mu\nu}+k^{\mu}k^{\nu})(1+\frac{\xi\beta^2}{2})+
\xi(k^2\beta^{\mu}\beta^{\nu}+(\beta\cdot k)^2\eta^{\mu\nu})-
2\Lambda\beta^{\mu}\beta^{\nu}\bigg]^{-1}.
\end{eqnarray}
The inverse matrices necessary to find these propagators can be found explicitly:
\begin{eqnarray}
(Q^{\mu}\gamma_{\mu}-M)^{-1}&=&(Q^{\nu}\gamma_{\nu}+M)\frac{1}{Q^2-M^2};\nonumber\\
\nonumber(A\eta^{\alpha\beta}+Bk^{\alpha}k^{\beta}+C \beta^{\alpha}\beta^{\beta})^{-1}&=&
\frac{1}{A}\eta_{\alpha\beta}-
\frac{(A+C\beta^2)B}{A\Delta}k_{\alpha}k_{\beta}-\\
&-&\frac{(A+Bk^2)C}{A\Delta}\beta_{\alpha}\beta_{\beta}+
\frac{BC}{A\Delta}(\beta\cdot k)(\beta_{\alpha}k_{\beta}+\beta_{\beta}k_{\alpha}),
\end{eqnarray}
where $\Delta=(A+Bk^2)(A+C\beta^2)-BC(\beta\cdot k)^2$. We note that for the case $\xi=0$ (absence of the non-metricity) this expression matches the result found in \cite{AltKost}.
In our case, $A=
-k^2(1+\frac{\xi\beta^2}{2})+\xi(\beta\cdot k)^2$, $B=1+\frac{\xi\beta^2}{2}$, $C=-2\Lambda+\xi k^2$. These expressions allow to write down our propagators:
\begin{eqnarray}
<\psi(-k)\bar{\psi}(k)>&=&i\bigg[\gamma^{\mu}[k_{\mu}(1-\frac{\xi}{4}\beta^2)-\frac{\xi}{2}\beta_{\mu}(\beta\cdot k)]+
m(1-\frac{\xi\beta^2}{2})\bigg]\\
\nonumber &\times&\bigg[k^2(1-\frac{\xi}{4}\beta^2)^2-\xi(1-\frac{\xi}{2}\beta^2)(\beta\cdot k)^2-
m^2(1-\frac{\xi\beta^2}{2})^2\bigg]^{-1}
;\nonumber\\
\nonumber<\tilde{B}_{\alpha}(-k)\tilde{B}_{\beta}(k)>&=&i\frac{1}{-k^2(1+\frac{\xi\beta^2}{2})+\xi(\beta\cdot k)^2}\Big[\eta_{\alpha\beta}-
\Delta^{-1}\Big([-k^2(1-\frac{1}{2}\xi\beta^2)+\xi(\beta\cdot k)^2-\\
&-&2\Lambda\beta^2](1+\frac{\xi\beta^2}{2})k_{\alpha}k_{\beta}+
\xi(\beta\cdot k)^2(-2\Lambda+\xi k^2)\beta_{\alpha}\beta_{\beta}-\nonumber\\\nonumber&-&
(1+\frac{\xi\beta^2}{2})
(-2\Lambda+\xi k^2)(\beta\cdot k)(\beta_{\alpha}k_{\beta}+\beta_{\beta}k_{\alpha})
\Big)\Big],
\end{eqnarray}
with
\begin{eqnarray}
\Delta&=&-(-2\Lambda+\xi k^2)(\beta\cdot k)^2(1+\frac{\xi\beta^2}{2})+\nonumber\\&+&
\xi(\beta\cdot k)^2[-k^2(1-\frac{1}{2}\xi\beta^2)-2\Lambda\beta^2+\xi(\beta\cdot k)^2].
\end{eqnarray}
We see that the exact propagator of the vector field is highly cumbersome.  However, the leading order in the two-point function of the vector field is of zero order in $\xi$, and of first order in the spinor field, since spinor-vector vertices already contain $\xi$.
Therefore, to consider the lower-order contributions, it is sufficient to take into account  in the propagator only zero order in $\xi$:
\begin{eqnarray}
\Delta&=&2\Lambda(\beta\cdot k)^2+O(\xi).
\end{eqnarray}
So, we can write our propagators as follows:
\begin{eqnarray}
\label{propzero}
<\psi(-k)\bar{\psi}(k)>&=&i\frac{\gamma^{\mu}k_{\mu}+m}{k^2-m^2}+O(\xi);\\
\nonumber<\tilde{B}_{\alpha}(-k)\tilde{B}_{\beta}(k)>&=&-i\frac{1}{k^2}\Big[\eta_{\alpha\beta}-
\frac{1}{2\Lambda(\beta\cdot k)^2}\Big([-k^2-2\Lambda\beta^2]k_{\alpha}k_{\beta}+\\
\nonumber&+&2\Lambda(\beta\cdot k)(\beta_{\alpha}k_{\beta}+\beta_{\beta}k_{\alpha})
\Big)\Big]
+O(\xi).
\end{eqnarray}
\begin{figure}[h]
	\centering
	\includegraphics[scale=1]{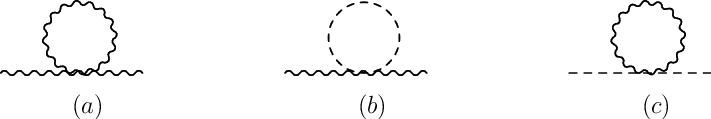}
	\caption{\label{Fig1} The contributions to the two-point functions.}
\end{figure}

With these propagators, we can find the contributions from graphs (a), (b), (c), and also of the graph (i), which we define later. It is worth  recalling that these graphs arise as a result of non-linear interactions produced by the non-metricity,  thus having no counterpart in the metric formulation of this theory.
As we already noted, we restrict ourselves to the first order in $\xi$ for external spinor legs, which means that since the spinor-vector vertex is already of the first order in $\xi$, for the graphs (b,c) we can disregard $\xi$-dependent terms in the denominator. As for the graph (a), we must keep first order in $\xi$ in the denominator when considering a contribution with $\lambda$ vertex, and disregard $\xi$ in the denominator when considering a contribution with $\xi$ vertex. As usual, the fermionic loop carries a minus sign. All that remains to do are just contractions. Unfortunately, the complexity of the complete propagator of the vector field that one finds is extremely cumbersome already at first order in $\xi$. For this reason, we only provide a qualitative description of possible contributions involving it, while the exact results will be presented only at zero order in $\xi$.

The graph (a) will contribute not only to the mass term but also to the kinetic term.  Both these contributions will diverge and, from the structure of the vertices, we conclude that we will have results proportional to $\frac{a\Lambda m^2+b\xi m^4}{\epsilon}\tilde{B}^{\mu}\tilde{B}_{\mu}$ (with $a,b$ some real numbers), to $\frac{\xi m^2}{\epsilon}(\beta^{\mu}B_{\mu})^2$, and to $\frac{\xi}{\epsilon}(\beta^{\mu}\tilde{B}_{\mu\nu})^2$, $\frac{\xi m^2}{\epsilon}\tilde{B}_{\mu\nu}\tilde{B}^{\mu\nu}$. Thus,  in general, we can obtain aether-like kinetic and mass terms. To obtain the explicit result, we consider only zero order in $\xi$ with the propagator (\ref{propzero}).
Thus, we have
\begin{eqnarray}
\Gamma_a&=&\frac{\Lambda}{4}(4\tilde{B}^{\alpha}\tilde{B}^{\beta}+2\eta^{\alpha\beta}\tilde{B}^{\gamma}\tilde{B}_{\gamma})
\int\frac{d^4k}{(2\pi)^4}\frac{1}{k^2}\Big[\eta_{\alpha\beta}-
\frac{1}{2\Lambda(\beta\cdot k)^2}\Big([-k^2-2\Lambda\beta^2]k_{\alpha}k_{\beta}-\nonumber\\
&-&2\Lambda(\beta\cdot k)(\beta_{\alpha}k_{\beta}+\beta_{\beta}k_{\alpha})
\Big)\Big]
+O(\xi) \ ,
\end{eqnarray}
which is formally highly singular. However, { by using that the tadpole integrals are identically zero, i.e.,
	\begin{eqnarray}
	\int \frac{d^{d}k}{(2\pi)^d}\frac{1}{k^2}=0, \,\, \int \frac{d^{d}k}{(2\pi)^d}\frac{1}{(\beta\cdot k)^2}=0\label{kk}
	\end{eqnarray}
	and 
	\begin{eqnarray}
	\int \frac{d^{d}k}{(2\pi)^d}\frac{1}{(\beta\cdot k)^n}=0,\,\, \mbox{with}\,\, n=0,1,2...\,\, ,
	\label{kkk}
	\end{eqnarray}
	together with the table of integrals given in \cite{Leib}, one can verify that $\Gamma_a$ vanishes.}

The graph (b) can be calculated explicitly at first order. Since we already have $\xi$ in the vertex, one can disregard any dependence of the spinor propagator on $\xi$ and leave the rest with the first order contribution in $\xi$:
\begin{eqnarray}
\nonumber\Gamma_b&=&-\frac{\xi}{2}{\rm tr}\int\frac{d^4k}{(2\pi)^4}\frac{1}{k^2-m^2}\bigg(-\gamma^{\mu}k_{\mu}\gamma_{\mu}k_{\nu}(\tilde{B}^{\mu}\tilde{B}^{\nu}+\frac{1}{2}B^2\eta^{\mu\nu})+
m^2B^2\bigg)\nonumber\\
&=&-\frac{\xi m^4}{16\pi^2\epsilon}\tilde{B}^{\mu}\tilde{B}_{\mu}+{\rm fin}.
\end{eqnarray}
This is just a simple renormalization of the mass term. The Lorentz-breaking vector $\beta_{\mu}$ will appear only in contributions of higher orders in $\xi$. 

For the graph (c) we see that the term in the propagator proportional to $\frac{1}{k^2}$ does not yield any contribution within the dimensional regularization framework. 
Actually we have
\begin{eqnarray}
\nonumber\Gamma_c&=&-i\frac{\xi}{2}\bar{\Psi}\Big[i(\delta^{\mu}_{\lambda}\delta^{\nu}_{\rho}+\frac{1}{2}\eta_{\lambda\rho}\eta^{\mu\nu})
\gamma_{\mu}\partial_{\nu}-m\eta_{\lambda\rho}\Big]\Psi\times\\
&\times&\int\frac{d^4k}{(2\pi)^4}\frac{1}{2(\beta\cdot k)^2}\Big([-k^2-2\Lambda\beta^2]k^{\lambda}k^{\rho}+
2\Lambda(\beta\cdot k)(\beta^{\lambda}k^{\rho}+\beta^{\rho}k^{\lambda})
\Big)\Big].
\end{eqnarray}

We see that this contribution superficially diverges. Similarly to the contribution $\Gamma_a$, one can use the formulas from the table of integrals \cite{Leib} to find that actually this integral disappears. Then,
we conclude that the contributions to the spinor sector are at least of second order in $\xi$ being thus strongly suppressed.

Moreover, in the case of a nontrivial vacuum there will be a new contribution to the two-point function of the vector field generated by the Feynman diagram with two triple vertices.

\begin{figure}[h]
	\centering
	\includegraphics[scale=1]{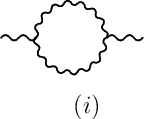}
	\caption{\label{Fig3} The new contribution to the two-point function.}
\end{figure}
Doing all contractions, we find that
\begin{eqnarray}
\nonumber\Gamma_i&=&\Lambda^2\beta_{\mu}\beta_{\nu}\bigg[\tilde{B}^{\mu}\tilde{B}^{\nu}<\tilde{B}^{\lambda}\tilde{B}^{\rho}><\tilde{B}_{\lambda}\tilde{B}_{\rho}>+
2\tilde{B}^{\lambda}\tilde{B}^{\rho}<\tilde{B}_{\lambda}\tilde{B}_{\rho}><\tilde{B}^{\mu}\tilde{B}^{\nu}>+\\
\nonumber&+&2\tilde{B}^{\lambda}\tilde{B}^{\rho}<\tilde{B}_{\lambda}\tilde{B}^{\nu}><\tilde{B}^{\mu}\tilde{B}_{\rho}>+
4\tilde{B}^{\lambda}\tilde{B}^{\nu}<\tilde{B}_{\lambda}\tilde{B}_{\rho}><\tilde{B}^{\mu}\tilde{B}^{\rho}>
\bigg]\\
&=&\Lambda^{2}\tilde{B}^{\mu}\tilde{B}^{\nu}\Pi_{\mu\nu},
\end{eqnarray}
where $\Pi_{\mu\nu}$ is conveniently broken into four contributions as below:  
\begin{eqnarray}
\Pi_{\mu\nu}=\Pi_{\mu\nu}^{(1)}+\Pi_{\mu\nu}^{(2)}+\Pi_{\mu\nu}^{(3)}+\Pi_{\mu\nu}^{(4)},
\label{pi}
\end{eqnarray}
with each contribution given by:
\begin{eqnarray}
\nonumber\Pi_{\mu\nu}^{(1)}&=&\beta_{\mu}\beta_{\nu}<\tilde{B}^{\lambda}\tilde{B}^{\beta}><\tilde{B}_{\lambda}\tilde{B}_{\beta}>\\
\nonumber&=&\beta_{\mu}\beta_{\nu}\int \frac{d^{d}k}{(2\pi)^{d}}\bigg(-\frac{1}{\Lambda(\beta\cdot k)}+\frac{2\beta^2}{k^{2}(\beta\cdot k)}-\frac{2}{k^4}-\\
\nonumber&-&\frac{k^4}{4\Lambda^{2}(\beta\cdot k)^4}+\frac{k^{2}\beta^{2}}{\Lambda(\beta\cdot k)^4}-\frac{\beta^4}{(\beta\cdot k)^4}-
\frac{2\beta^2}{k^{2}(\beta\cdot k)^2}\bigg);
\end{eqnarray}
\begin{eqnarray} 
\nonumber\Pi_{\mu\nu}^{(2)}&=&2\beta_{\alpha}\beta_{\beta}<\tilde{B}_{\mu}\tilde{B}_{\nu}><\tilde{B}^{\alpha}\tilde{B}^{\beta}>\\
\nonumber&=&\int \frac{d^{d}k}{(2\pi)^{d}}\bigg(\frac{4\beta^{2}(\beta_{\mu}k_{\nu}+\beta_{\nu}k_{\mu})}{k^{4}(\beta\cdot k)}+\frac{(\beta_{\mu}k_{\nu}-\beta_{\nu}k_{\mu})}{\Lambda k^{2}(\beta\cdot k)}+
\frac{\eta_{\mu\nu}}{\Lambda k^2}+\frac{\beta^{2}k_{\mu}k_{\nu}}{\Lambda k^{2}(\beta\cdot k)}\bigg);
\end{eqnarray}
\begin{eqnarray}
\nonumber\Pi_{\mu\nu}^{(3)}&=&2\beta_{\alpha}\beta_{\beta}<\tilde{B}_{\mu}\tilde{B}^{\beta}><\tilde{B}^{\alpha}\tilde{B}_{\nu}>\\
\nonumber&=&\int \frac{d^{d}k}{(2\pi)^{d}}\bigg(-\frac{\beta^{2}(k_{\mu}\beta_{\nu}+k_{\nu}\beta_{\mu})}{\Lambda k^{4}(\beta\cdot k)}-\frac{\beta^{2}k_{\mu}k_{\nu}}{\Lambda k^{2}(\beta\cdot k)^4}-\\
\nonumber&-&\frac{2\beta^{4}2k_{\mu}k_{\nu}}{\Lambda k^{4}(\beta\cdot k)^4}+\frac{\beta^{2}(k_{\mu}\beta_{\nu}-k_{\nu}\beta_{\mu})}{\Lambda k^{4}(\beta\cdot k)}+\\
\nonumber&+&\frac{(\beta_{\mu}k_{\nu}+\beta_{\nu}k_{\mu})}{\Lambda k^{2}(\beta\cdot k)}-\frac{2\beta^{2}(\beta_{\mu}k_{\nu}+\beta_{\nu}k_{\mu})}{k^{4}(\beta\cdot k)}-
\frac{k_{\mu}k_{\nu}}{2\Lambda^{2}(\beta\cdot k)^2}-\frac{4\beta^{4}k_{\mu}k_{\nu}}{k^{4}(\beta\cdot k)}
\bigg);
\end{eqnarray}
\begin{eqnarray}
\nonumber\Pi_{\mu\nu}^{(4)}&=&4\beta_{\alpha}\beta_{\nu}<\tilde{B}_{\mu}\tilde{B}_{\lambda}><\tilde{B}^{\alpha}\tilde{B}^{\lambda}>\\
\nonumber&=&\int \frac{d^{d}k}{(2\pi)^{d}}\left(-\frac{2\beta_{\mu}\beta_{\nu}}{\Lambda(\beta\cdot k)^2}-\frac{k^{2}k_{\mu}\beta_{\nu}}{\Lambda^{2}(\beta\cdot k)^3}+2\frac{\beta^{2}k_{\mu}\beta_{\nu}}{\Lambda(\beta\cdot k)^3}\right).
\end{eqnarray}
We can use the table of integrals in Appendix A of \cite{Leib} as well as the formulas (\ref{kk}) and (\ref{kkk}) to calculate the above contributions and conclude that all of them vanish altogether. Such a result was to be expected because the external momenta are identically zero.  Therefore, the diagram (i) does not contribute to the effective potential. Taking into account also the vanishing of the zero-order contribution from the diagram (a), we conclude that the effective potential is at least of  first order in $\xi$, like the contribution (b). In fact, the complete effective potential within certain prescriptions is obtained under the suggestion that the vector field is purely external while the spinors are completely integrated out, in such a way that (b) is the only non-trivial lower contribution to the effective potential.     




Before concluding, let us give here a brief discussion of possible contributions to the four-point functions. While their full analysis is much more involved than in the trivial vacuum scenario, we note that in our case contributions like $(B^{\mu}B_{\mu})^2$ and $B^{\mu}B_{\mu}\bar{\Psi}\Psi$ will also arise just as for the trivial vacuum. In addition, it is natural to expect terms like $(\beta^{\mu}B_{\mu})^2B^2$ and $(\beta^{\mu}B_{\mu})^2\bar{\Psi}\Psi$, which however must be at least of second order in $\xi$. In whole analogy with the two-point functions, it is natural to expect that nontrivial contributions to the four-point function will also arise at least at first order in $\xi$. Accordingly, the effect of the non-metricity will be important in the first and higher orders in $\xi$.

\section{Summary and conclusions}\label{sec4}

In this work we formulated the metric-affine bumblebee model coupled to a spin 1/2 field and computed the corresponding quantum corrections in the weak gravitational field limit up to $\mathcal{O}(\xi^2,\lambda\xi,\lambda^2)$. We employed a perturbative approach  within the framework of the effective field theory methodology, where the role of the inverse square of the energy scale is played by the nonminimal coupling constant $\xi$. In this regard, we discussed the small $\xi$ dominant contributions (zero and first orders in $\xi$) to two- and four-point functions of all fields, spinor and vector ones, finding that the theory is renormalizable up to this order\footnote{Obviously, the whole theory including gravity is affected by the same renormalizability problems as standard GR.}. These calculations have been performed for the non-trivial vacuum, where one has a constant vector $\beta^{\mu}$ that gives rise to a highly unusual propagator, only observed previously in very specific studies of confinement \cite{Leib}. However, many of the lower-order contributions to the effective action simply vanish, so that the effective potential  includes terms at least of first order in $\xi$, which allows us to conclude that quantum corrections associated to the non-minimal coupling, $\xi B^{\mu}B^{\nu}R_{\mu\nu}$, are highly suppressed in comparison with the classical action. Moreover, we observe that the spinor sector is even more suppressed, involving the terms that begin at second order in $\xi$ or higher. We expect that higher-order  quantum contributions in this theory can be used to explore specific phenomenological aspects. Also, we note that the effective potential of $B_{\mu}$, obtained up to the fourth order in the vector field, can be used to explore the possibility of dynamical Lorentz symmetry breaking. These issues are better understood by using the Schwinger-DeWitt method which is the more appropriate approach to explore quantum corrections to the effective potential in metric-based theories. It would be interesting to check their viability in theories defined in non-Riemaniann backgrounds that is the case of the metric-affine bumblebee model. These analyses,  together with aspects of gravitatinal waves and non-perturbative classical solutions, are currently under way.

\acknowledgments

This work was partially supported by Conselho Nacional de Desenvolvimento Cient\'{\i}fico e Tecnol\'{o}gico (CNPq), Coordena\c{c}\~{a}o de Aperfei\c{c}oamento de Pessoal de N\'{\i}vel Superior (CAPES), by the Spanish Grant FIS2017-84440-C2-1-P funded by  MCIN/AEI/10.13039/501100011033 “ERDF Away of making Europe”, Grant PID2020-116567GB-C21 funded by MCIN/AEI/10.13039/501100011033, the project  PROMETEO/2020/079 (Generalitat Valenciana), and the project i-COOPB20462 (CSIC).  The work by A. Yu. P. has been supported by the CNPq project No. 301562/2019-9. PJP would like to thank the Brazilian agency CAPES for financial support (PNPD/CAPES grant, process 88887.464556/2019-00) and Departament de F\'{i}sica Te\`{o}rica and IFIC, Universitat de Val\`{e}ncia, for the hospitality. A. D. is supported by a PhD contract of the program FPU 2015 (Spanish Ministry of Economy and Competitiveness) with reference FPU15/054/06, and would like to thank the Departamento de F\'{\i}sica, Universidade Federal da Para\'{\i}ba for hospitality.



\end{document}